\documentclass[%
 aip,
 amsmath,amssymb,floatfix,
 reprint,%
]{revtex4-1}

\usepackage{graphicx}
\usepackage{dcolumn}
\usepackage{bm}

\usepackage[utf8]{inputenc}
\usepackage[T1]{fontenc}
\usepackage{mathptmx}
\usepackage{url}

\begin{document}

\preprint{AIP/123-QED}

\title[Magnetoresistive Sensor Detectivity]{Magnetoresistive Sensor Detectivity: A Comparative Analysis}

\author{J. E. Davies}
 \email{jdavies@nve.com}
 \author{J. D. Watts}
 \author{J. Novotny}
 \author{D. Huang}
 \author{P. G. Eames}

\affiliation{NVE Corporation, Eden Prairie, MN 55344, USA}

\date{\today}

\begin{abstract}
We report on the noise performance characteristics of magnetic sensors using both magnetic tunnel junction (MTJ) and giant magnetoresistance (GMR) elements. Each sensor studied has a notably different noise and detectivity. Of the sensors we measured, those based on GMR multilayers have the lowest noise and detectivity. However, the GMR sensor also has a significantly smaller linear range. To make a direct comparison between sensors we scale the linear operating ranges of each sensor to be the same. This is the phenomenological equivalent of modifying the flux concentration. Upon scaling the low frequency detectivity of the TMR sensors becomes essentially equal to that of the GMR sensor.  Using the scaling approach we are able to place the detectivity in the context of other key parameters, namely size and power consumption.  Lastly, we use this technique to examine the upper limit for magnetoresistive sensor performance based on a notional MTJ sensor using present record setting TMR values.  
\end{abstract}

\maketitle

Magnetoresistive (MR) technologies have been the fundamental building block for spintronic devices over the last three decades.\cite{Baibich1988,Binasch1989,Dieny1991,Parkin1999} This is due to their ability to serve as highly sensitive transducers that readily respond to magnetic fields and spin currents.\cite{slon2002,Ralph2008,sankey2008,Wang1997,Ikeda2008,Wang2009,Wolf2001} Their small size and high sensitivity are utilized as sensor components across several industries including the biomedical, navigation and industrial automation markets.\cite{GRAH2004,caru1997,SCHE1997,Fujiwara2018} They are also at the core of the magnetic data storage industry, being utilized as read heads in hard disk drives and the storage medium for magnetic random access memory (MRAM) bits.\cite{McFadyen2006,Parkin1999} From the sensor perspective, the MR sensor's CMOS compatible fabrication process, robust performance, low power operation and low cost are clear advantages over other technologies. These advantages poise MR sensors to play a key role in technologies benefiting from advanced, compact, and cost effective sensing, such as the internet of things (IoT), smart grids and electric cars. \cite{Tahoori2018,XLiu2019,OUYANG20198}

With many types of MR sensors available it is often difficult to determine the best sensor for a particular application. Typically, application engineers have used maximization of the \% MR and the minimization of noise as key metrics.\cite{Nowak1998,Ikeda2008,STUT2005}  These metrics are straightforward for the researcher to focus on with much work having gone into the development of MR sensor noise phenomenology.\cite{Nowak1998,PARK2000,KLAS2004,Zheng2019} Experimental noise characterization has shown that there can be substantial variation in the noise characteristics among sensors.\cite{STUT2005,EGEL2009,DEAK17} It is important to realize that such comparisons between specific sensors can be misleading in that they do not place the comparison in the context of other important specifications such as sensor size, hysteresis, power consumption, linear range and cost. Performing this more rigorous contextual comparison is non-trivial, but more telling. 

This work places magnetoresistive sensor noise characterization in a more parameterized context. We start with a noise and performance study of three sensor variations. Each sample has a different field response and noise spectrum. Normalizing the noise data to a key parameter, namely the linear range, we obtain a more contextual picture of each sensor's performance and demonstrate that such normalization enables a more fundamental comparison between different sensors. Finally, we look ahead to what is likely the ultimate limit of MR sensor detectivity.

\begin{figure}
\includegraphics[width=0.3\textwidth]{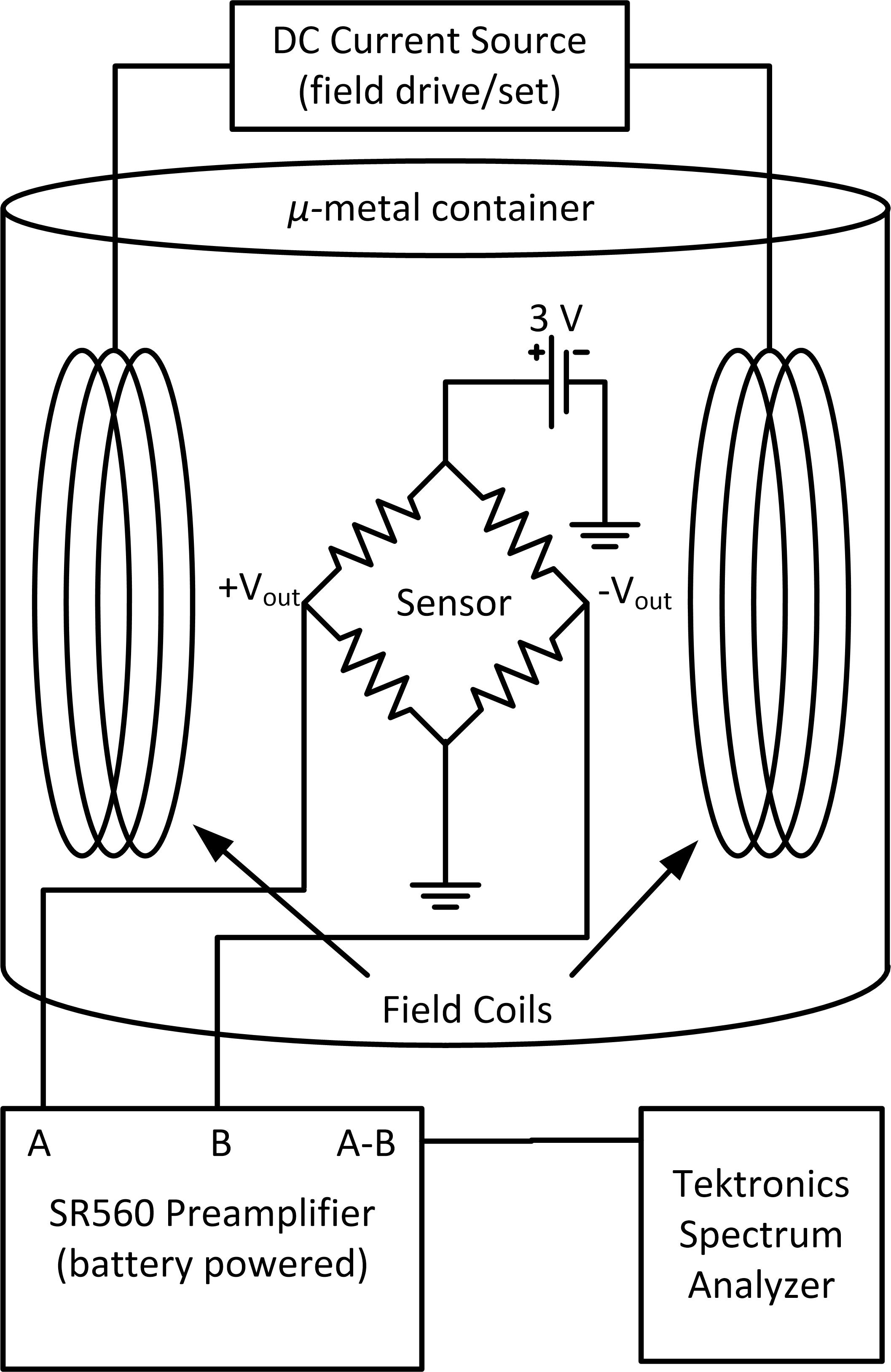}
\caption{\label{fig:setup} Illustration of the noise measurement setup. The sensor resides in a mu-metal can and is battery powered to 3 V. Helmholtz coils are available for field biasing of the sensor. Sensor outputs are fed to a low-noise preamplifier. The differential signal is then captured on a spectrum Analyzer.}
\end{figure}

The noise measurement setup is shown in Fig.\ref{fig:setup}. The sensors, configured as Wheatstone bridges, are soldered to circuit boards and placed in a triple layered $\mu$-metal container for shielding from environmental magnetic fields. The sensors were all battery powered to 3 V and, in the case of the unipolar sensor, a 0.5 mT field was applied along its sense direction to put the sensor in the middle of its operate range. The outputs of the bridges, $+V_{out}$ and $-V_{out}$ were input to a two-channel preamplifier. The differential output was then captured by a spectrum analyzer.

We limit this study to three often utilized sensor compositions (layer thicknesses are in nm):

\begin{enumerate}
  \item Bipolar MTJ sensor with superparamagnetic freelayer (SPMTJ) with layer composition: Ta(5)/Ru(5)/IrMn(11)/CoFe(2)/Ru(0.9)/CoFeB(2.5)/ MgO(2)/CoFeB(1.2)/Ta(2)/Ru(10)
  \item Bipolar MTJ sensor with full film free layer (FFMTJ) with layer composition: Ta(5)/Ru(5)/IrMn(11)/CoFe(2)/Ru(0.9)/CoFeB(2.5)/ MgO(2)/CoFeB(2)/NiFe(6)/Ta(2)/Ru(10)
  \item Omnipolar GMR multilayer sensor\cite{ZOU2001} (GMRMLs) with layers grown on a Ta(5) seed and comprised of four CoFe(1)/NiFeCo(2)/CoFe(1) ferromagnetic trilayers separated by CuAgAu(1.5) GMR spacers and capped with Ta(20)
\end{enumerate}

All films were grown by magnetron sputtering onto Si wafers coated with 200 nm vapor deposited SiNx. The SPMTJ and FFMTJ films were annealed ex-situ in a 400 mT field at 350C in a forming gas environment to both crystalize the MgO layer and set the pinning layer magnetization. No annealing was performed on the GMRML films. 

The sensor elements were fabricated using standard photolithography processes. SPMTJ and FFMTJ resistor elements are series connected arrays of forty 5 $\mu$m diameter MTJs. Individual resistors were then rotated and co-packaged to form a push-pull sensor configuration. 

The GMRML sensor was monlithically fabricated with four serpentine-style resistors. Thick NiFe layers were then used in dual purpose to shield two resistors and provide flux concentration for the active resistors.

\begin{table}
\caption{\label{tab:table1}Nominal sensor parameters}
\begin{ruledtabular}
\begin{tabular}{cccccc}
 Sensor & MR &$R_0$ &$B_{sat}$ &$Sens_{max}$\footnotemark[1] &Area\footnotemark[2]\\
 Type & (\%) &$(k\Omega)$ &(mT) &(mV/V/mT) &($mm^{2}$)\\
\hline
SPMTJ & 60 & 50 & 5 &25 &6.25\\
FFMTJ & 120 & 10 & 15 &25 &6.25\\
GMRML & 15 & 5 & 1 &12 &20\\
\end{tabular}
\end{ruledtabular}
\footnotetext[1]{1 mV/V/mT = 0.1 mV/V/Oe}
\footnotetext[2]{Area is circuit board area occupied by the sensor. SPMTJ and FFMTJ sensors are in TDFN-6 packages, GMRML is in SOIC-8.}
\end{table}

\begin{figure}
\includegraphics[width=0.45\textwidth]{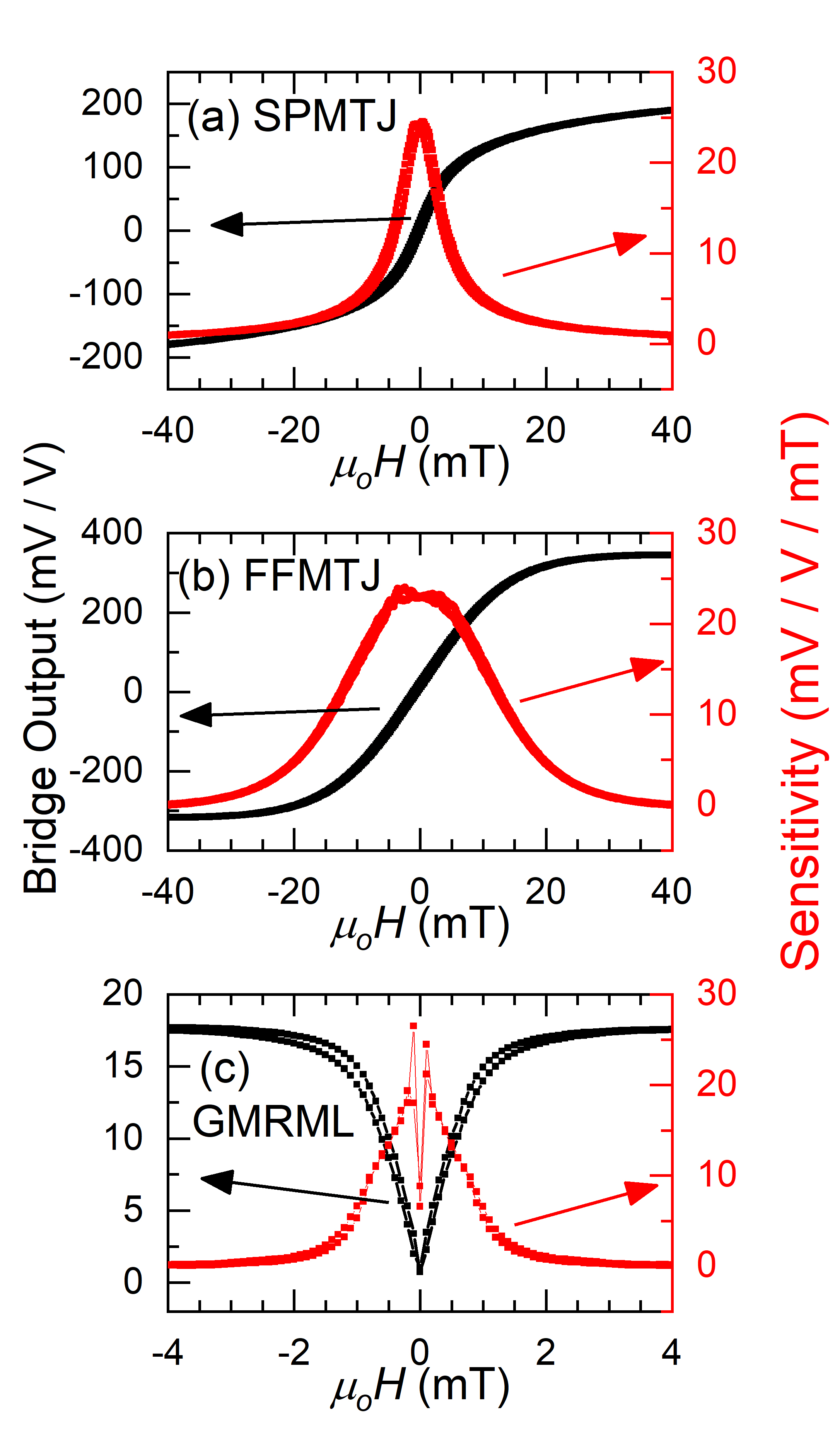}
\caption{\label{fig:curves} (left axes, black symbols) Bridge outputs and (right axes, red symbols) sensitivities for (a) SPMTJ sensor, (b) FFMTJ and (c) GMRML sensors. Noise measurements were performed and detectivity evaluated at \textit{H} =  0 mT for the SPMTJ and FFMTJ and \textit{H} = 0.5 mT for the GMRML.}
\end{figure}

The TMR ratio, bridge resistance (at $\mu_oH$ = 0 mT), saturation field ($B_{sat}$) and sensitivity for each sensor are listed in Table \ref{tab:table1}. Sensor outputs and sensitivities as a function of applied field for the three sensors are shown in Fig. \ref{fig:curves} as black and red curves, respectively.

Fig. \ref{fig:curves}a shows the typical performance of SPMTJ sensors. The bridge output has near zero hysteresis due to the thermal demagnetizing effects of the superparamagnetic free layer. There is no offset in the output. This is due to the fast randomization of the free layer's magnetization and granular structure of the film negating any of the traditional coupling effects. The peak sensitivity of 25 mV/V/mT (Fig. \ref{fig:curves}a, red) occurs at $H$ = 0 mT. The sensitivity decreases precipitously to either side of $H$ = 0 mT. The full width at half maximum (FWHM) of the sensitivity "peak" is approximately 10 mT. The sharpness of the sensitivity peak limits the linearity to +/- 2 mT.

The FFMTJ sensor response is shown in Fig. \ref{fig:curves}b. The FFMTJ was tailored to minimize hysteresis and offset while extending the linear range through the use of shape anisotropy. Interestingly, the maximum sensitivity is comparable to the SPMTJ sensor (Fig.\ref{fig:curves}b, red), however the sensitivity "peak" is much broader with a FWHM of 30 mT. This results in the linear range of the sensor being +/-10 mT, a factor of five increase over the SPMTJ films while preserving the sensitivity.

The GMRML sensor response (Fig. \ref{fig:curves}c) is exotic compared to the bipolar films. The multilayer structure results in a unipolar sensor response. Permalloy flux concentrators are used in order to maximize the sensitivity and tailor the linear range without modifying the underlying film properties. In this case, the flux concentrators employed allow ~10x reduction in the $B_{sat}$ (note the field axis compared to the MTJ sensors). The flux concentration allows the maximum sensitivity (Fig. \ref{fig:curves}c, red) to be comparable to the MTJ sensors (\ref{tab:table1}). This peak sensitivity occurs near $H$ = 0 mT. Without a bias field or magnet in place, the omnipolar device produces a discontinuity in the bridge output at $H$ = 0 mT. The usable range of the unbiased sensor shown is between 0.1 mT and 1.5 mT, which has an average sensitivity of 12 mV/V/T.

\begin{figure}
\includegraphics[]{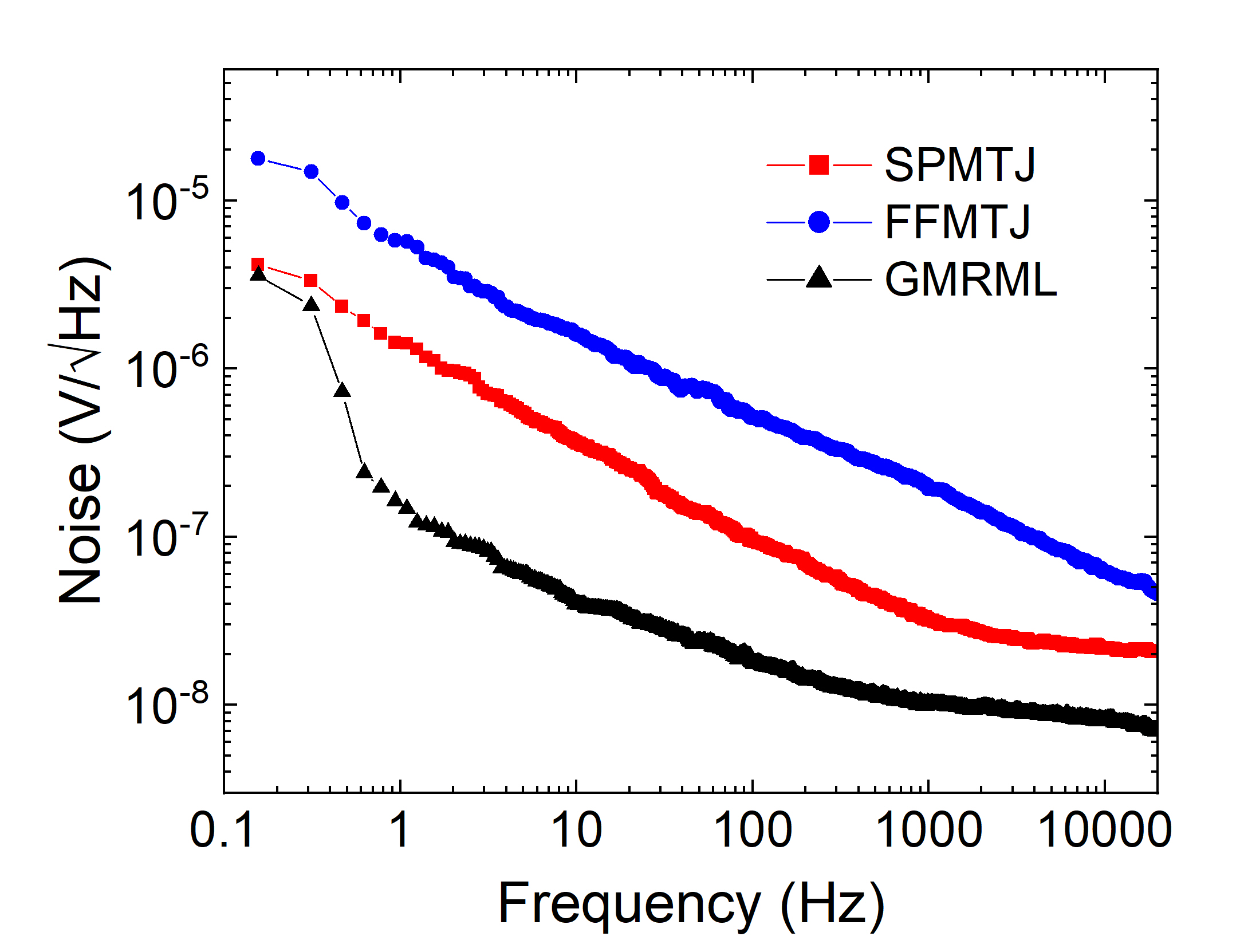}
\caption{\label{fig:Noise} Noise spectra for the GMRML (black), SPMTJ (red) and FFMTJ (blue) sensor elements in Fig. \ref{fig:curves}. Each sensor was biased to 3 V for the measurement as illustrated in Fig. \ref{fig:setup}.}
\end{figure}

Noise data from for the three sensors at their average operating fields (0 mT for the SPMTJ and FFMTJ and 0.5 mT for the GMRML) is shown in Fig. \ref{fig:Noise}. At low frequencies the noise is dominated by the magnetic $1/f$ contribution. The SPMTJ (Fig. \ref{fig:Noise}, red) and FFMTJ (Fig. \ref{fig:Noise}, blue) sensors have a similar $1/f$ character (slope). However, the SPMTJ sensor is roughly 10x less noisy. The GMRML sensor's spectrum (Fig. \ref{fig:Noise}, black) has a different character with more curvature and overall lower noise throughout the frequency range.

Around 1 kHz, the SPMTJ spectra (Fig. \ref{fig:Noise}, red) flattens out as the output noise approaches the Johnson noise floor. The frequency at which this transition takes place is increased for lower resistance devices. Thus, by 20 kHz the difference in noise between the high resistance SPMTJ and the FFMTJ has been reduced, and at still higher frequencies, the noise of the SPMTJ likely surpasses that of the FFMTJ.

The GMRML sensor also begins to saturate at high frequency, but does so more gradually (Fig. \ref{fig:Noise}, black). Previous noise measurements of this particular make of GMRML sensor (an NVE Corp. AA002) by Stutzke, \textit{et al.} show  the noise floor is reached around 100 Hz. The GMRML sensor from this study appears to have pushed the floor to higher frequencies and is likely due to our sensors being biased to 3 V compared to the 1.2 V in the Stutzke study, resulting in a larger electric $1/f$ contribution.\cite{STUT2005,JIAN2004}

\begin{figure}
\includegraphics[]{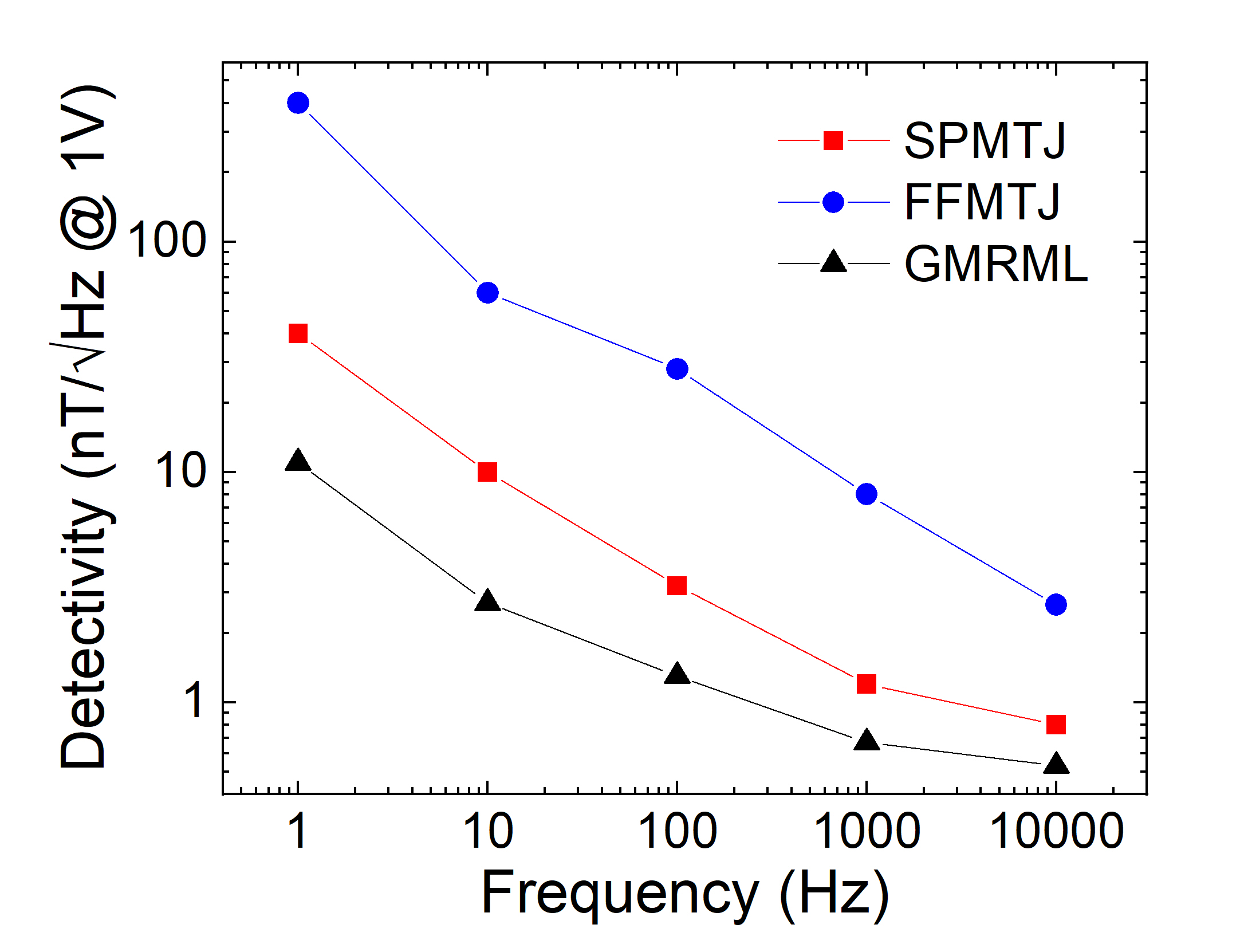}
\caption{\label{fig:Detectivity} Detectivity versus frequency for the GMRML (black), SPMTJ (red)and FFMTJ (blue) sensor elements in Fig. \ref{fig:curves} The 2.5x difference between the SPMTJ and GMRML sensors allows for the SPMTJ sensor to have the lowest detectivity at 10 kHz.}
\end{figure}

Dividing the noise spectra by the sensitivity yields the detectivity, i.e. the lowest detectable field. This is shown in Fig. \ref{fig:Detectivity} for each sensor. The GMRML sensor (Fig. \ref{fig:Detectivity}, black triangles) has the lowest detectivity. The SPMTJ film has the second lowest detectivity; still ~67\% larger than the GMRML's throughout the frequency range. The detectivity for both the SPMTJ and GMRMR drops below 1 nT/$\sqrt{Hz}$ at 10 kHz with values of 0.8 nT/$\sqrt{Hz}$ and 0.53 nT/$\sqrt{Hz}$, for the SPMTJ and GMRML sensors, respectively. 

It is important to note that while the GMRML sensor has the lowest detectivity, it also has the smallest linear range with $B_{sat}$ = 1 mT. According to the $B_{sat}$ values in Table I, a flux concentration of 5x and 15x for the SPMTJ and FFMTJ films, respectively, would result in sensors with the same linear range. This provides for a more direct comparison of the sensors.

\begin{figure}
\includegraphics[]{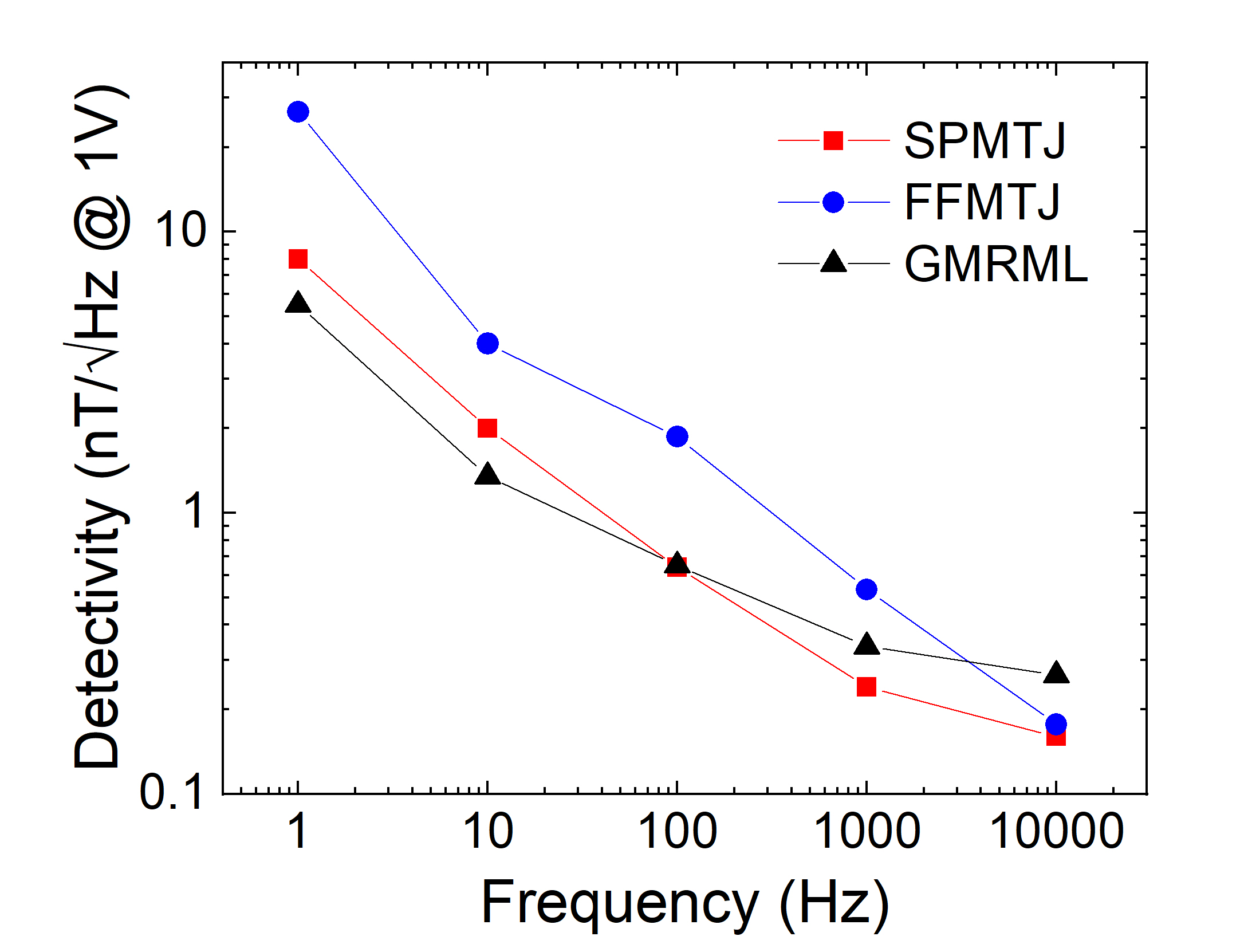}
\caption{\label{fig:DetectivityLinrange} Detectivity versus frequency for the GMRML (black), SPMTJ (red) and FFMTJ (blue) sensor elements. This time the detectivity curves are normalized to have the same linear range. The normalization allows for a more direct comparison of the sensor materials.}
\end{figure}

Fig. \ref{fig:DetectivityLinrange} shows the detectivity of each of the three sensors as if they were flux concentrated to have the same $B_{sat}$. The GMRML and SPMTJ films now have comparable detectivities of 6 nT/$\sqrt{Hz}$  and 8 nT/$\sqrt{Hz}$ at 1 Hz, respectively. The SPMTJ detectivity drops below that of the GMRML for $f$  > 100 Hz. In contrast, the FFMTJ still has a detectivity of 30 nT/$\sqrt{Hz}$ at 1 Hz, only becoming comparable to the other sensors above 1 kHz.

This comparative analysis relies on the ability to manipulate the flux concentration to create equivalent sensors. In practice, adding flux concentration results in two main performance trade-offs. First, even modest amounts of flux concentration (e.g. 5x or higher) will drastically increase the sensor size. Second, adding flux concentration  reduces and limits the operating field range of the sensor. 

The second trade-off presents the dilemma for magnetoresistive sensors of choosing detectivity minimization versus having a robust operating field range. This is an important problem to address in applications such as surgical navigation and precision automation where fields on the order of 1 mT are used, but sub 1 nT/$\sqrt{Hz}$ resolution are required. Surgical navigation sensors also need to be very small; ideally less than 2 mm in any direction.

\begin{figure}
\includegraphics[]{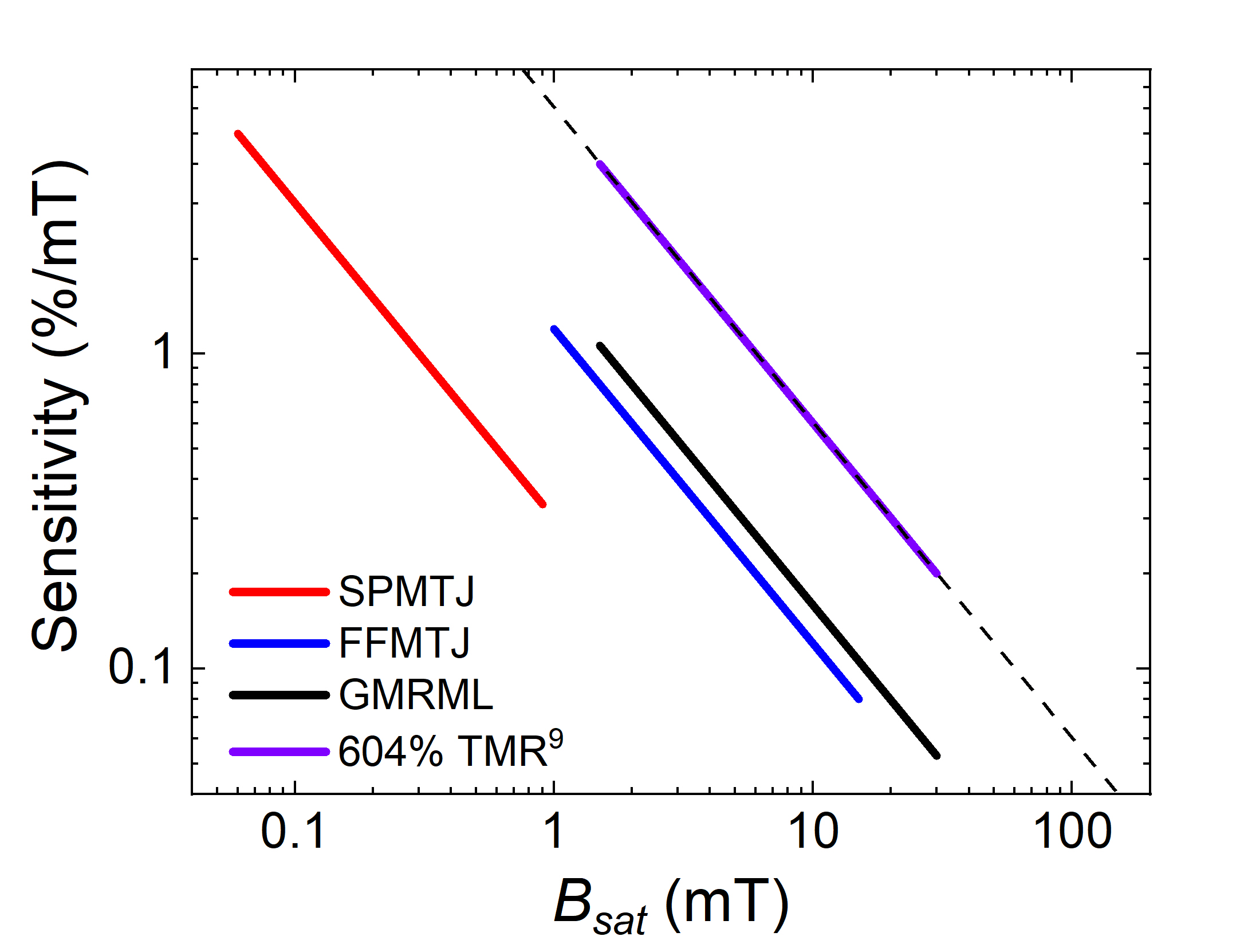}
\caption{\label{fig:Ikeda} Plot of sensitivity versus $B_{sat}$ for the (red) SPMTJ, (blue) FFMTJ and (black) GMRML sensors. (purple) A notional sensor, assuming the ability to create a device with record MgO TMR experimentally achieved by Ikeda, \textit{et al.} is also shown\cite{Ikeda2008,Zheng2019}. Lines take each sensor type from 1x to 20x flux concentration. The coincident dashed line represents the potential ultimate limit of TMR sensor performance.}
\end{figure}

One option to address such requirements is to maximize the TMR ratio.  It has been shown that TMR ratios beyond 100 \% won't significantly improve sensor noise, but will allow for a reduced need for flux concentration, and hence a higher linear range.\cite{EGEL2009} Fig. \ref{fig:Ikeda} shows the possible linear ranges, i.e. $(B_{sat})$ versus sensitivities for various GMR and TMR sensors. Each line shows when the sensor is flux concentrated between 1x (highest $B_{sat}$ value) and 20x (lowest $B_{sat}$ value). Sensors at higher $B_{sat}$ and sensitivity correlate with larger TMR ratios. 

The record room temperature TMR ratio is 604\%, demonstrated by Ikeda \textit{et al.}\cite{Ikeda2008} The long standing of this record makes it something of an upper limit for MTJ devices. That particular MTJ is not practical for a sensor for several reasons, most specifically it being in a pseudo-spin-valve configuration and having large hysteresis. However, it can notionally be used to show the limits of TMR sensor detectivity. Thus, included in Fig. \ref{fig:Ikeda} is a hypothetical sensor with 604\% TMR. The dashed line is used to delineate the ultimate limits of TMR sensitivity versus $B_{sat}$. Comparing measured literature values from the MR Sensor roadmap, it appears that MR sensor detectivity will minimize around 10 pT/$\sqrt{Hz}$ at 1 Hz without significant material changes or technology shifts.\cite{Zheng2019}

\begin{figure}
\includegraphics[]{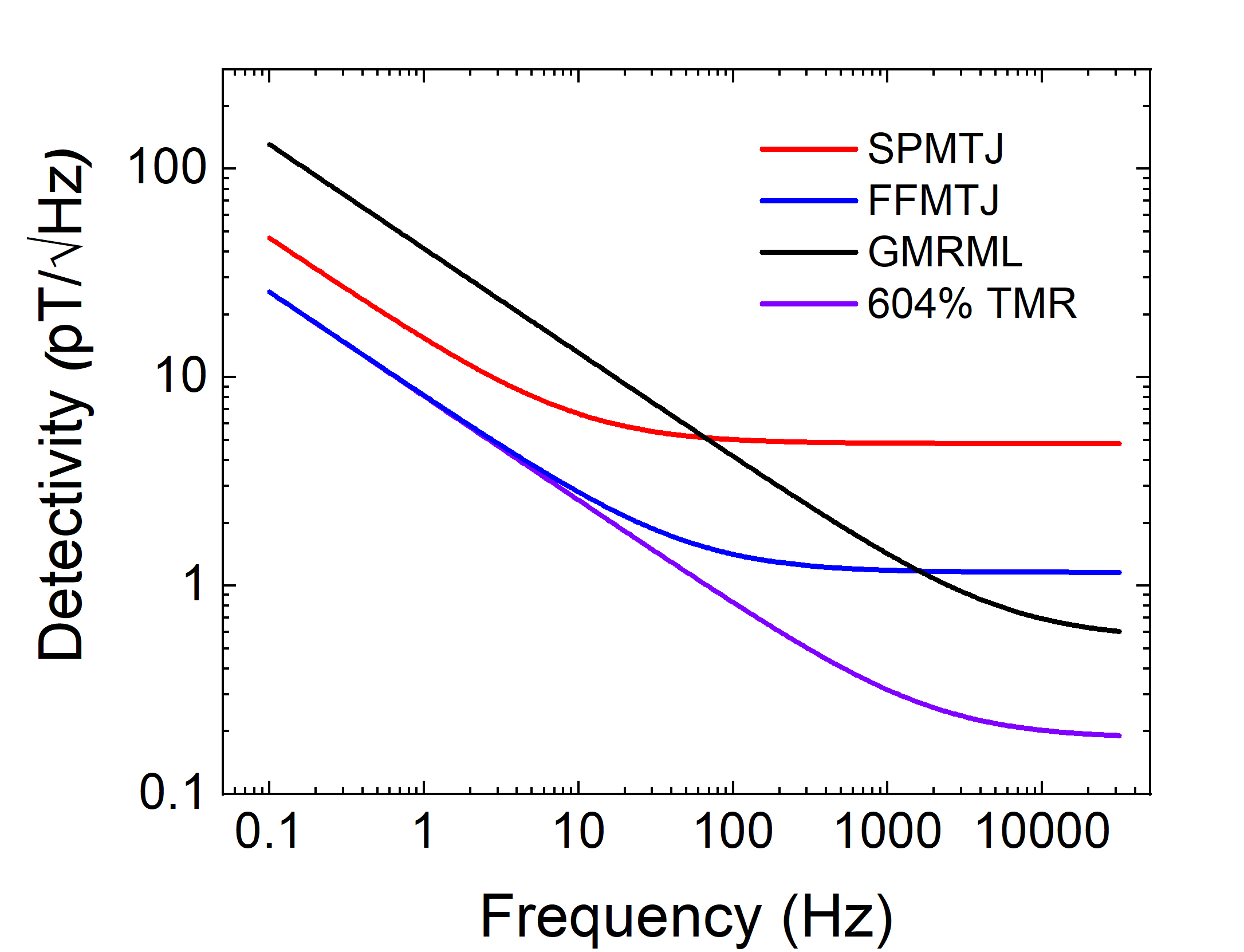}
\caption{\label{fig:Ideal} Modeled ultimate detectivity versus frequency for the (red) SPMTJ, (blue) FFMTJ, (black) and (purple) notional sensor reaching 604\% TMR.\cite{Ikeda2008} All four plots assume $B_{sat}$ = 0.1 mT, no additional flux concentration and an active resistor area of 3 mm$^2$.}
\end{figure}

Fig. \ref{fig:Ideal} shows models of the detectivity versus frequency for the SPMTJ, FFMTJ, GMRML and notional film. The models constrain parameters such that the sensor footprint and magnetic response are as identical as possible. We assume an "earth's field anomaly" application where $B_{sat}$ = 0.1 mT and there is a reasonably small component with an active resistor area of 3 mm$^2$. The small $B_{sat}$ and large resistor area allows for a significant drop in the detectivity compared to the measured values in Fig. \ref{fig:Detectivity}.

At low frequencies (1 Hz), where $1/f$ noise dominates, the detectivity of the MTJ materials is comparable with the FFMTJ and notional films having the lowest detectivity at roughly 10 pT/$\sqrt{Hz}$. This reaffirms the diminishing impact of TMR > 100 \% as shown by other groups. \cite{EGEL2009}. The SPMTJ sensor has the next lowest detectivity of ~20 pT/$\sqrt{Hz}$. This performance degradation is primarily attributed to the lower TMR ratio, although this could also be hampered by active area (which is generally an inaccessible parameter to an end user). The GMRML sensor has the highest detectivity at 70 pT/$\sqrt{Hz}$.

As is known, the $1/f$ contribution subsides at high frequencies, leaving Johnson and shot noise as the remaining detectivity contributions. These are dictated by the device resistance. With a RA product of nearly 400 k$\Omega-\mu$m$^2$ The FFMTJ film approaches 1 pT/$\sqrt{Hz}$ at ~1 kHz. The GMRML film's low resistance results in 700 fT/$\sqrt{Hz}$ detectivity around 100 kHz. The Ikeda film, with an RA = 10 $\Omega-\mu$m$^2$ allows for the detectivity to drop to 200 fT/$\sqrt{Hz}$ at 30 kHz.

Much work involving reaching fT detectivities involves some form of field or signal modulation. Of these are microelectromechanical system-based (MEMS) flux concentrators that serve to modulate the field at high frequencies where Johnson noise is the only consideration.\cite{EDEL06}. Another approach is to use specialized flux concentrators.\cite{Pann2006}. Indeed, the utilization of other effects such as modulation by voltage controlled magnetic anisotropy or spin transfer torques may also help to drop the detectivity.

Interestingly, it should be noted that going to high frequencies can result in an \textit{increased} detectivity.  Recent work by He \textit{et al.} has shown that the detectivity in flux concentrated MTJs can actually increase for large frequencies, as the permeability of the flux concentrators decrease; limiting the minimum detectivity in their sensor to  30 pT/$\sqrt{Hz}$.\cite{HE2019}  Thus, the technique of sensor/field modulation may also be limited to a particular frequency.

In conclusion we have performed a comparative study of the noise and detectivity in three classes of sensors. We have found that on first inspection, the GMRML sensor has the lowest detectivity, it also has the largest size and power consumption. When the sensors are normalized to $B_{sat}$ the differences are no longer evident. Extending the parameterized comparisons to other materials, including the best demonstration Ikeda films, shows that there is likely an ultimate performance limit for magnetoresistive sensors in the 100s of fT range. However, normalizing the sensor's noise performance to a tunable parameter, such as the linear operating range provides a much clearer and useful comparison. The hope is this work will serve as a guide to MR sensor design, illustrating the present limits of the technology.

We would like to acknowledge Cathy Nordman and Maria Torija for fruitful discussions. 

The data that support the findings of this study are available from the corresponding author upon reasonable request.

\bibliography{Bibliography.bib}

\end{document}